\documentclass[11pt,twoside]{article}
\usepackage{./asp2014}

\aspSuppressVolSlug
\resetcounters

\begin{document}

\title{Cold gas in High-z Galaxies: The dense ISM}
\author{R.~Decarli,$^{1,2}$ C.~Carilli$^{3,4}$, C.~Casey$^5$, B.~Emonts$^6$, J.A.~Hodge$^7$, K.~Kohno$^8$, D.~Narayanan$^9$, D.~Riechers$^{10}$, M.T.~Sargent$^{11}$, F.~Walter$^{2,3}$
\affil{$^1$INAF -- Osservatorio di Astrofisica e Scienza dello Spazio, via Gobetti 93/3, 40129 Bologna, Italy; \email{roberto.decarli@inaf.it}}
\affil{$^2$Max Planck Institut f\"{u}r Astronomie, K\"{o}nigstuhl 17, 69117 Heidelberg, Germany}
\affil{$^3$National Radio Astronomy Observatory, Pete V.~Domenici Array Science Center, P.O.~Box O, Socorro, NM 87801, USA}
\affil{$^4$Cavendish Laboratory, University of Cambridge, 19 J.J.~Thomson Avenue, Cambridge CB3 0HE, UK}
\affil{$^5$Department of Astronomy, The University of Texas at Austin, 2515 Speedway Blvd Stop C1400, Austin, TX 78712}
\affil{$^6$National Radio Astronomy Observatory, 520 Edgemont Road, Charlottesville, VA 22903, USA}
\affil{$^7$Leiden Observatory, Niels Bohrweg 2, 2333 CA Leiden, The Netherlands}
\affil{$^8$Institute of Astronomy, School of Science, The University of Tokyo, 2-21-1 Osawa, Mitaka, Tokyo 181-0015, Japan}
\affil{$^9$Department of Astronomy, University of Florida, 211 Bryant Space Science Center, Gainesville, FL 32611, USA}
\affil{$^{10}$Cornell University, 220 Space Sciences Building, Ithaca, NY 14853, USA}
\affil{$^{11}$Astronomy Centre, Department of Physics and Astronomy, University of Sussex, Brighton, BN1 9QH, UK}
}

\paperauthor{R.~Decarli}{roberto.decarli@inaf.it}{0000-0002-2662-8803}{INAF}{Osservatorio di Astrofisica e Scienza dello Spazio}{Bologna}{BO}{40129}{Italy}
\paperauthor{C.~Carilli}{ccarilli@nrao.edu}{}{NRAO}{Pete V.~Dominici Array Science Center}{Socorro}{NM}{87801}{USA}
\paperauthor{C.~Casey}{cmcasey@astro.as.utexas.edu}{}{University of Texas at Austin}{Department of Astronomy}{Austin}{TX}{78712}{USA}
\paperauthor{B.~Emonts}{bjornemonts@gmail.com}{}{National Radio Astronomy Observatory}{50 Edgemont Road}{Charlottesville}{VA}{22903}{USA}
\paperauthor{J.A.~Hodge}{hodge@strw.leidenuniv.nl}{}{Leiden Observatory}{}{Leiden}{CA}{2333}{The Netherlands}
\paperauthor{K.~Kohno}{kkohno@ioa.s.u-tokyo.ac.jp}{}{School of Science, University of Tokyo}{Institute of Astronomy}{Tokyo}{Mitaka}{181-0015}{Japan}
\paperauthor{D.~Narayanan}{desika.narayanan@gmail.com}{}{University of FLorida}{Department of Astronomy}{Gainesville}{FL}{32611}{USA}
\paperauthor{D.~Riechers}{riechers@astro.cornell.edu}{}{Cornell University}{}{Ithaca}{NY}{14853}{USA}
\paperauthor{M.T.~Sargent}{Mark.Sargent@sussex.ac.uk}{}{University of Sussex}{Astronomy Centre, Department of Physics and Astronomy}{Brighton}{}{BN1 9QH}{UK}
\paperauthor{F.~Walter}{walter@mpia.de}{}{MPIA}{Galaxies and Cosmology}{Heidelberg}{}{69117}{Germany}


\section{Science Goals}
The goal of this science case is to study physical conditions of the interstellar medium (ISM) in distant galaxies. In particular, its densest component is associated with the inner cores of clouds -- this is where star formation takes place. Carbon monoxide is usually used to trace molecular gas emission; however, its transitions are practically opaque, thus preventing astronomers from piercing through the clouds, into the deepest layers that are most intimately connected with the formation of stars. Other dense gas tracers are required, although they are typically too faint and/or at too low frequencies to be effectively observed in high redshift galaxies. The ngVLA will offer for the first time the sensitivity at radio frequencies that is needed to target [C{\sc i}]$_{1-0}$ (at $z>5$), as well as the ground transitions of dense gas tracers of the ISM such as HCN, HNC, HCO+ (at various redshifts $z>1$), beyond the tip of the iceberg of the hyper-luminous sources that could be studied up to now. These new tools will critically contribute to our understanding of the intimate interplay between gas clouds and star formation in different environments and cosmic epochs.


\section{Scientific rationale}

While undoubtedly the carbon monoxide (CO) rotational emission lines are extremely useful to gauge the molecular gas mass in distant galaxies, they have a limited power in exposing the diversity of properties and the complexity of the multi-phased ISM. For instance, both observational and theoretical studies revealed substantial amount of diffuse molecular gas that is not associated with CO emission \citep[e.g.,][]{langer14,glover16}. On the other hand, CO transitions reach optical depth values close to unity even at very modest column densities, i.e., CO-emitting regions appear opaque. Furthermore, the critical density required to collisionally de-excite CO is relatively low ($n_{\rm H2}\sim 10^2-10^3$ cm$^{-3}$ for the lower-J transitions), meaning that CO is a fairly poor tracer of the dense molecular cores, where star-formation within distant galaxies is ultimately taking place.

The neutral carbon fine structure lines [C{\sc i}] at 609 and 370 $\mu$m arise in the external parts of molecular clouds, where the radiation field is too intense for CO molecules to survive, but at not enough to ionize the carbon atoms (optical extinction $A_V=1-5$ mag). The 370 $\mu$m transition has been detected even at $z\approx 7$ \citep{venemans17}. The two neutral carbon fine structure lines are ideal tracers of the more diffuse molecular ISM \citep[see, e.g.,][]{glover15}. In particular, because of the different critical densities, their ratio is an excellent tracer of the gas density. Furthermore, in combination with other lines (such as CO or [C{\sc ii}]) they can be used to infer physical properties of the ISM such as the gas density and the intensity of the incident radiation field \citep[see, e.g.,][]{meijerink08,popping18}. [C{\sc i}] is also considered a complementary tracer of the cold gas mass, which can be used to infer either an independent measurement of the CO--to--H$_2$ conversion factor, $\alpha_{\rm CO}$, or to measure the neutral carbon abundance, $X_{\rm CI}$ \citep[see, e.g.,][]{bothwell17,popping17}. 

To date, only about 40 galaxies at $z>1$ have been detected in [C{\sc i}]. This list mostly consists of sub-mm galaxies and quasar host galaxies. E.g., \citet{bothwell17} use [C{\sc i}] 609\,$\mu$m observations, in combination with other ISM lines, to infer molecular gas mass, density and intensity of the radiation field in 13 dusty, star--forming galaxies at 2$<${}$z${}$<$5 using ALMA. They find a higher density and stronger UV field for the star--forming medium in these galaxies than estimated by previous studies of similar sources. At higher redshifts, however, the [C{\sc i}]$_{1-0}$ transition is shifted outside the ALMA bands at $z>4.8$. This ground transition is a better mass tracer than the [C{\sc i}]$_{2-1}$ transition (observable with ALMA at these high redshifts), which is also sensitive to the excitation temperature. Furthermore, only the detection of both these lines provides a direct measure of the neutral gas density. The detection of the [C{\sc i}]$_{1-0}$ line at higher redshift thus requires sensitive observations in the frequency ranges that will be covered by the ngVLA. 

In addition to neutral carbon, other molecular gas tracers can provide us with precious complementary information than CO. E.g., high dipole moment molecules like hydrogen cyanide (HCN) are only collisionally de-excited at very high densities, making them much more reliable tracers of the very dense gas directly associated with the formation of individual stars. Some studies of HCN in the nearby universe have even found evidence that the ratio of HCN luminosity to FIR luminosity remains constant over $>$8 orders of magnitude in HCN luminosity, suggesting that HCN may be a fundamental direct probe of star forming `units', and that the only difference between star formation on different scales and in different environments is the number of these fundamental star-forming units \citep[e.g.,][]{gao04,wu05,wu10,zhang14}. 

However, not much is known about dense gas tracers in the more distant universe. Owing to the fact that they only trace the densest regions of the ISM and are therefore less abundant, emission from the rotational transitions of molecules like HCN is usually an order of magnitude fainter than CO, complicating efforts to detect and study these tracers at high-$z$. As a result, only a few high-$z$ galaxies have been detected in dense gas tracers to-date --- most of which are either intrinsically extremely luminous, or strongly lensed \citep[e.g.,][]{solomon03,vandenbout04,carilli05,gao07,riechers07,riechers11,danielson11}. With  ALMA  now  online,  the  situation  will clearly  improve dramatically in the near future. However, as with CO, ALMA will only be able to  detect the mid- and high-J transitions of dense gas tracers like HCN, HNC, and HCO+. These higher-level transitions are less directly tied to the total dense gas mass, requiring assumptions about the (highly uncertain) excitation ratios. In addition, these higher-J transitions are more likely to be affected by IR pumping, which local studies find may be common in ultraluminous galaxies \citep[e.g.,][]{aalto07}. Thus, while necessary to understand the overall excitation properties of high-$z$ sources, the ALMA-detectable transitions may be unsuitable as tracers of the total dense gas mass. 

\begin{figure}
\centering
\includegraphics[width=0.99\textwidth]{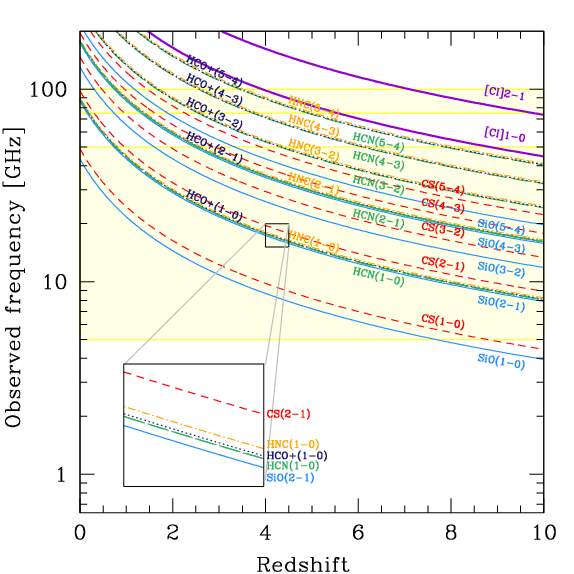}
\caption{Redshifted frequencies of dense gas tracers, many of which will be accessible in single-tuning setups with the ngVLA at high-redshift. In particular, the 1-0 and 2-1 transitions of HNC, HCO+, HCN and SiO lines are spaced very closely in frequency (see inset zoom-in). While ALMA can detect the mid- and high-J transitions of these molecules at high-redshift, significant uncertainties regarding excitation mean that the low-J transitions accessible with the ngVLA are more suitable as tracers of the total dense gas mass. }
\label{fig_wg3_dense_freqrange}
\end{figure}

The current JVLA probes the right frequency range to detect the crucial low-J transitions of these high critical density molecules (Figure \ref{fig_wg3_dense_freqrange}). However, its limited sensitivity means that the current state-of-the-art for high-$z$ detections consists of a smattering of strongly lensed hyper-starbursting quasar hosts \citep[e.g.,][]{vandenbout04,riechers07,riechers11}. With significantly increased sensitivity, the ngVLA would extend studies of the dense gas mass at high-$z$ beyond this handful of extreme objects for the first time. In addition to tracing the dense gas mass at high-$z$, such studies are critical for constraining models of star formation based on the gas density PDF, as these models make testable predictions about the FIR-HCN relation in FIR-luminous objects \citep[e.g.,][]{krumholz07,narayanan08}. Finally, while angular resolution is not a priority for these photon-starved studies, the brightest objects could even be spatially resolved by a ngVLA on $\sim$kpc scales (requiring baselines on the order of the current VLA). The ngVLA would thus enable detailed studies of the dense gas at high-redshift such as are currently only possible in the local universe.

\begin{figure}
\centering
\includegraphics[width=0.99\textwidth]{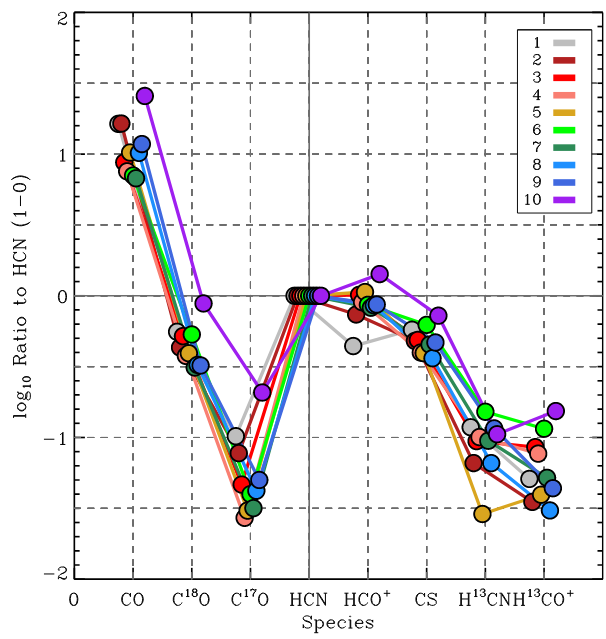}
\caption{The intensity (in units of temperature) of the ground transitions of CO, C$^{18}$O, C$^{17}$O, HCN, HCO$^+$, CS, and H$^{13}$CNH$^{13}$CO$^+$ in various clouds (labeled 1--10) in the local starburst galaxy NGC253, normalized to the observed HCN(1-0) flux. CO is at least 10$\times$ brighter than other dense gas tracers -- in order to detect these other lines at high redshift, the unprecedented sensitivity of ngVLA at radio frequencies will be instrumental. Figure adapted from \citet{leroy15}.}
\label{fig_wg3_dense_leroy15}
\end{figure}

\subsection{Measurements Required}

This science goal requires integrated flux measurements of faint lines in high redshift galaxies. The sensitivity and the coverage in the radio bands are prime technical requirements for these observations. The targeted lines will be $>$10$\times$ fainter than CO (in temperature units). A galaxy with stellar mass $M_{\rm *}\approx 10^{11}$\,M$_\odot$ on the massive main sequence of star--forming galaxies at $z\sim2$ has SFR$\approx$100\,M$_\odot$\,yr$^{-1}$ \citep{santini17}. This corresponds to an IR luminosity $L_{\rm IR}\approx 5\times 10^{11}$\,L$_\odot$. Assuming the CO--to--IR luminosity empirical relation from \citet{carilli13}, we infer an expected CO(1-0) luminosity of $7.0\times10^9$ K\,km\,s$^{-1}$\,pc$^2$, or 345,000 L$_\odot$. For a CO(1-0)/HCN(1-0)$\approx$10 (see Fig.~\ref{fig_wg3_dense_leroy15}), we infer an expected flux density for the HCN(1-0) transition of 7\,$\mu$Jy (assuming a line width of 300\,km\,s$^{-1}$). Such a line can be detected at 3-$\sigma$ significance with 26 hr of integration, assuming that ngVLA will have 5$\times$ the collecting area of the current JVLA. 
No existing instrument has the sensitivity at radio frequencies to perform the proposed observations.

Imagining requirements are limited at $\sim1-3''$ resolution, in order to avoid confusion with other sources in the field, and at the same time limit the risk of out-resolving the faint emission. A large bandwidth would allow to simultaneously cover multiple ISM tracers (see Figure \ref{fig_wg3_dense_freqrange}) in one shot, i.e., it would be ideal although it is not a strict requirement.


\subsection{Longevity/Durability: with respect to existing and planned ($>$2025) facilities}

There are obvious synergies with other key observatories to envision in the context of the characterization of the ISM in high-$z$ galaxies. In particular, ALMA can provide coverage of the intermediate and high-J transitions of the dense gas tracers, as well as of the [C{\sc i}]$_{2-1}$ line, whereas ngVLA will focus on the lower-J lines and on the [C{\sc i}]$_{1-0}$ transition (see Fig.~\ref{fig_wg3_dense_freqrange}). The combinations of CO transitions ranging from low to high J, together with the denser gas tracers targeted with ngVLA + ALMA, will enable an unprecedentedly detailed study of the properties of the densest parts of the ISM, which are not currently accessible. This will enable a direct link between the gas properties and the on-going star formation, given that the dense gas tracers are intimately connected with birthplace of stars. JWST and the 30m class telescope will provide sensitive spectroscopy of rest-frame UV/optical/NIR ISM tracers, mostly associated with the ionized phase of the gas, thus paving the way for a coherent description of the gas cycle in galaxies throughout cosmic history.

\acknowledgements ...  



\end{document}